\documentclass[12pt]{iopart}

\usepackage{iopams}  
\usepackage{graphicx} 
\newcommand{\Benard}{B\'enard }

\begin{document}

\title{Two-Particle Dispersion in Weakly Turbulent Thermal Convection}

\author{S Sch\"utz$^{1,2}$\footnote{Author to whom any correspondence should be addressed} and E Bodenschatz$^{2}$}

\address{$^1$ Emergent Complexity in Physical Systems (ECPS), \'Ecole Polytechnique F\'ed\'erale de Lausanne (EPFL), Station 9, 1015 Lausanne, Switzerland}
\address{$^2$ Max Planck Institute for Dynamics and Self-Organization (MPIDS), Am Fa{\ss}berg 17, 37077 G\"ottingen, Germany}
\eads{\mailto{simon.schuetz@epfl.ch}, \mailto{eberhard.bodenschatz@ds.mpg.de}}
\begin{abstract}
We present results from a numerical study of particle dispersion in the weakly nonlinear regime of Rayleigh-\Benard convection of a fluid with Prandtl number around unity, where bi-stability between ideal straight convection rolls and weak turbulence in the form of Spiral Defect Chaos exists. While Lagrangian pair statistics has become a common tool for studying fully developed turbulent flows at high Reynolds numbers, we show that key characteristics of mass transport can also be found in convection systems that show no or weak turbulence. Specifically, for short times, we find  Richardson-like $t^3$ scaling of pair dispersion.  We explain our findings quantitatively with a model that captures the interplay of advection and diffusion. For long times we observe diffusion-like dispersion of particles that becomes independent of the individual particles' stochastic movements. The spreading rate is found to depend on the degree of spatio-temporal chaos. 
\end{abstract}

\submitto{\NJP}
\maketitle

% Broken: TOC title overwrites paper title
\tableofcontents

\section{Introduction}
A defining characteristic of fluid flow is its ability to transport mass \cite{Toschi.2009.AnnuRevFM,Salazar.2009.AnnuRevFM}.
One well-established description of transport is the dispersion of fluid elements from an initial separation $r_0$, i.e. the study of their mean square separation $\langle r^2\rangle$ with time.

A substantive description of transport has been derived for fully developed, homogeneous, and isotropic turbulence that has a well developed inertial range \cite{Salazar.2009.AnnuRevFM}.
In many types of such turbulent systems \cite{Salazar.2009.AnnuRevFM,Fung.1998.PRE,Eyink.2011.PhysRevE}, dispersion has been shown to be Richardson-like \cite{Obukhov1941,Batchelor1950,Richardson.1926.PRSL} in the inertial range (i.e. $\langle\Delta^2\rangle\sim t^3$).
However, flows in nature show spatio-temporal chaos or weak turbulence that do not display the scale invariance of fully developed hydrodynamic turbulence.
 
Here, we report numerical and analytical findings for Lagrangian dispersion of diffusing passive tracers in a vertically strongly confined flow system at very low Reynolds numbers. 
For two-particle dispersion, we find analogies with fully developed turbulence and trace them back to simple arguments that are valid for both systems.
The system we studied is large aspect ratio thermal convection of an incompressible Prandtl number one fluid,  where we can clearly differentiate between effects at length scales smaller than the vertical length scale, and the influence of the vertical confinement that affects horizontal transport for larger distances.

\section{Methods}

\subsection{Vertically Confined Laminar and Weakly Turbulent Flow}
\begin{figure}[!t]
\centering
\includegraphics[scale=0.8]{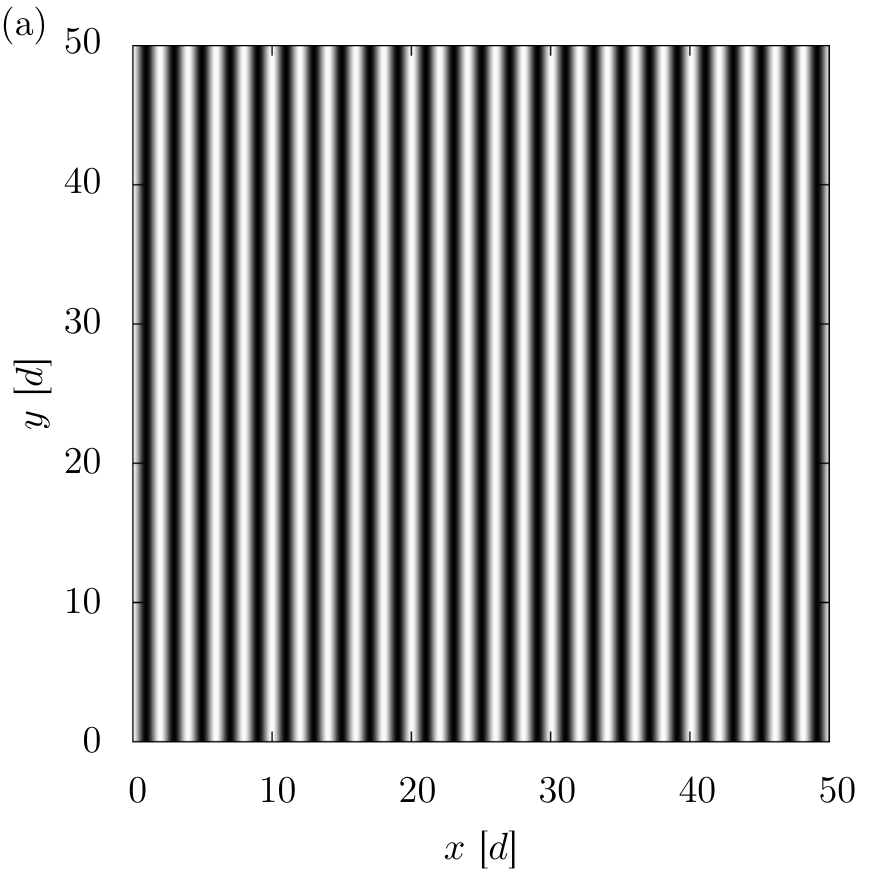}
\includegraphics[scale=0.8]{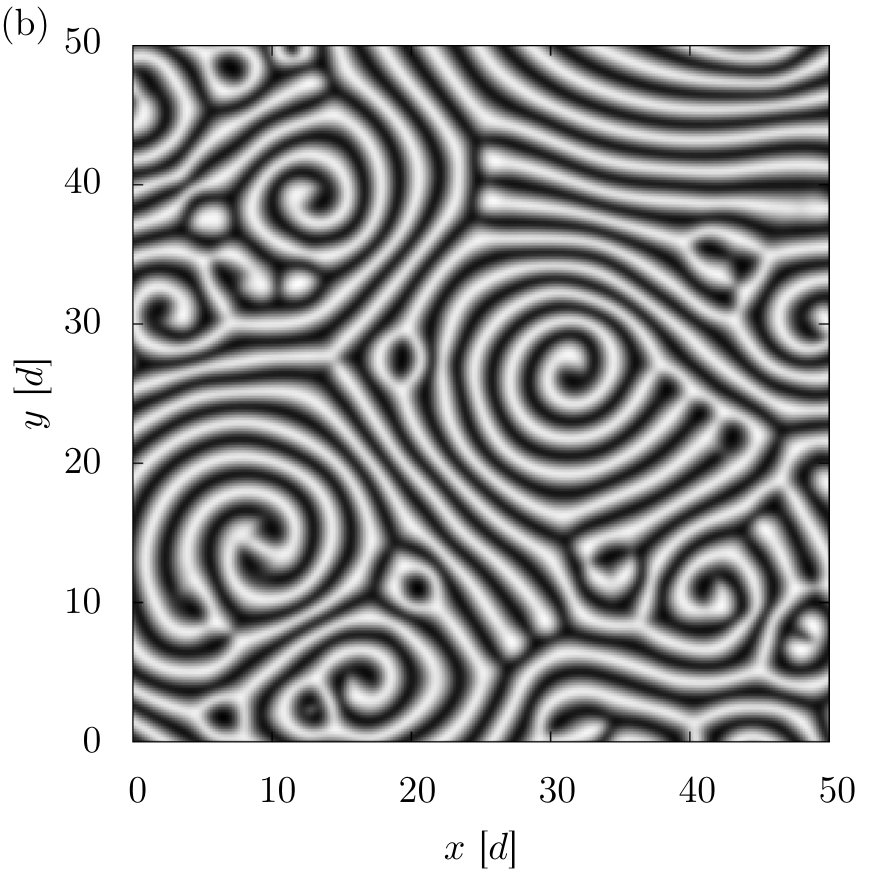}
\caption{Bistable convection patterns of Rayleigh-\Benard convection at $Pr=1$, visualized by the temperature at the midplane between top and bottom plate. Darker color marks a lower temperature. Only part of the domain is visible, the full system fills a periodic domain of size $L=100\,d$. (a) Straight Roll pattern (here at $Ra=2647$), common for low Rayleigh numbers close to convection onset. (b) Spiral Defect Chaos (here at $Ra=3074$), common at higher Rayleigh numbers. Numerical results from our simulation.}
\label{fig:midplanetemp}
\end{figure}
Consider a large aspect ratio Rayleigh-\Benard system, where a thin layer fluid is contained between two horizontal plates which are held at different temperatures.
For a Newtonian fluid of kinematic viscosity $\nu$ and thermal diffusivity $\kappa$, the convective dynamics of the system depends on two dimensionless quantities, which are the Prandtl number $Pr$ and Rayleigh number $Ra$, with
\begin{equation}
Pr\equiv\nu/\kappa \qquad\textrm{and}\qquad Ra\equiv\alpha g d^3 \Delta T /(\nu\kappa).
\end{equation}
Here, $\alpha$ is the volume expansion coefficient $\alpha$, $g$ is the magnitude of gravitational acceleration, $d$ the vertical distance between the plates and $\Delta T$ is the temperature difference between the plates (where a positive $\Delta T$ corresponds to the lower plate being warmer, i.e. an unstable fluid layering).
Onset of convection occurs at the critical Rayleigh number $Ra_c\approx1708$.
Closely above onset, the flow forms convection rolls \cite{Bodenschatz.2000.AnnRevFM} that, depending on initial conditions, align in parallel or display a time-dependent state known as Spiral Defect Chaos \cite{Cakmur.1997.PRL}.

The \Benard system allows us to realize, in the same geometry, two qualitatively different flow patterns to study transport.
The straight roll pattern has a discrete and a continuous translational symmetry in the two horizontal directions and is largely time independent, which makes the flow pattern easy to describe.
However, many systems in nature do not have that high a degree of regularity.
The periodicity and translation invariance can be broken by studying the same system in the state of Spiral Defect Chaos. 

For our study, we numerically solved the incompressible Navier-Stokes equations and heat equation under the Boussinesq approximation \cite{Boussinesq.1897.Book} in a 2D-periodic domain of edge length $L=100\,d$ or $L=200\,d$ (with $d$ the vertical system size).
Straight rolls or spiral defects were selected by imposing infinitesimal temperature fluctuations to the initial conditions.
The equations were discretized using a pseudospectral Galerkin scheme originally developed by Werner Pesch, and each flow component expressed at a resolution between $1024\times1024\times2$ and $2048\times2048\times5$ spectral basis functions.
Time integration used an implicit-explicit scheme of time step $\rmd t=10^{-2}$.
Details of the GPU-based parallel numerical solver can be found in \ref{sec:solver_details}.

\subsection{Dispersion of Diffusing Passive Tracers}
In this flow, we followed the trajectories of passive massless particles which, in addition to being advected with the fluid velocity $\bi{u}$, experienced an individual Brownian diffusion with diffusion coefficient $\mathcal{D}$.
For a small, but finite time interval $\rmd t$, a particle's position $\bi{x}$ is described by the Langevin-like advection-diffusion relation
\begin{eqnarray}
\rmd\bi{x}/\rmd t=\bi{u}(\bi{x},t)+\bfeta(t)/\rmd t,
\end{eqnarray}
where $\bfeta(t)$ is a random jump direction drawn from a Gaussian distribution of standard deviation $\sigma=\sqrt{6\mathcal{D} \rmd t}$ with no temporal correlation. 

Particles were advected under a 4th-order Runge-Kutta time stepping scheme, where the particle velocity $\bi{u}_i$ was retrieved via trilinear interpolation from sampling the spectral flow field at a resolution of $1024\times1024\times128$, making use of the fact that the flow field $\bi{u}(\bi{x},t)$ changed on a time scale much larger than the advection time scale.
The random diffusion jump $\bfeta(t)$ was added to the particle position after each time step $\rmd t$.
The total number of advected particles was between $10^3$ and $10^5$ for each simulation run.

We considered the pair dispersion $R^2(t)$, defined as
\begin{eqnarray}
R^2(t) \equiv \langle r^2_{ij}(t)\rangle
\end{eqnarray}
where $r_{ij}\equiv |\bi{x}_j-\bi{x}_i|$, and $\langle\ldots\rangle$ denotes an averaging over a representative ensemble of particle pairs $i$, $j$ with $r_{ij}(t_0)=0$.
Note here that the presence of diffusion allows us to set the initial separation to zero, whereas nondiffusing tracers need a small initial separation $0<r_{ij}(t_0)\ll1$ for their deterministic trajectories to ever separate.

Unless noted otherwise, we measure distances in terms of the system height $d$ and particle diffusivities in terms of the thermal diffusivity $\kappa$.
As a velocity scale, we take the average flow speed in the system,
\begin{equation}
U\equiv \langle |\bi{u}(\bi{x})|\rangle_\Omega \label{eq:U},
\end{equation}
where $\langle\ldots\rangle_\Omega$ denotes a spatial average over the volume $\Omega$.
This velocity scale depends on the flow configuration, i.e. both the external parameters (such as the Rayleigh number), and the flow pattern.
The flow velocity and system height define the vertical advection time $d/U$, which we take as our time scale.

\section{Results}
\begin{figure}[b]
\centering
\includegraphics[scale=0.8]{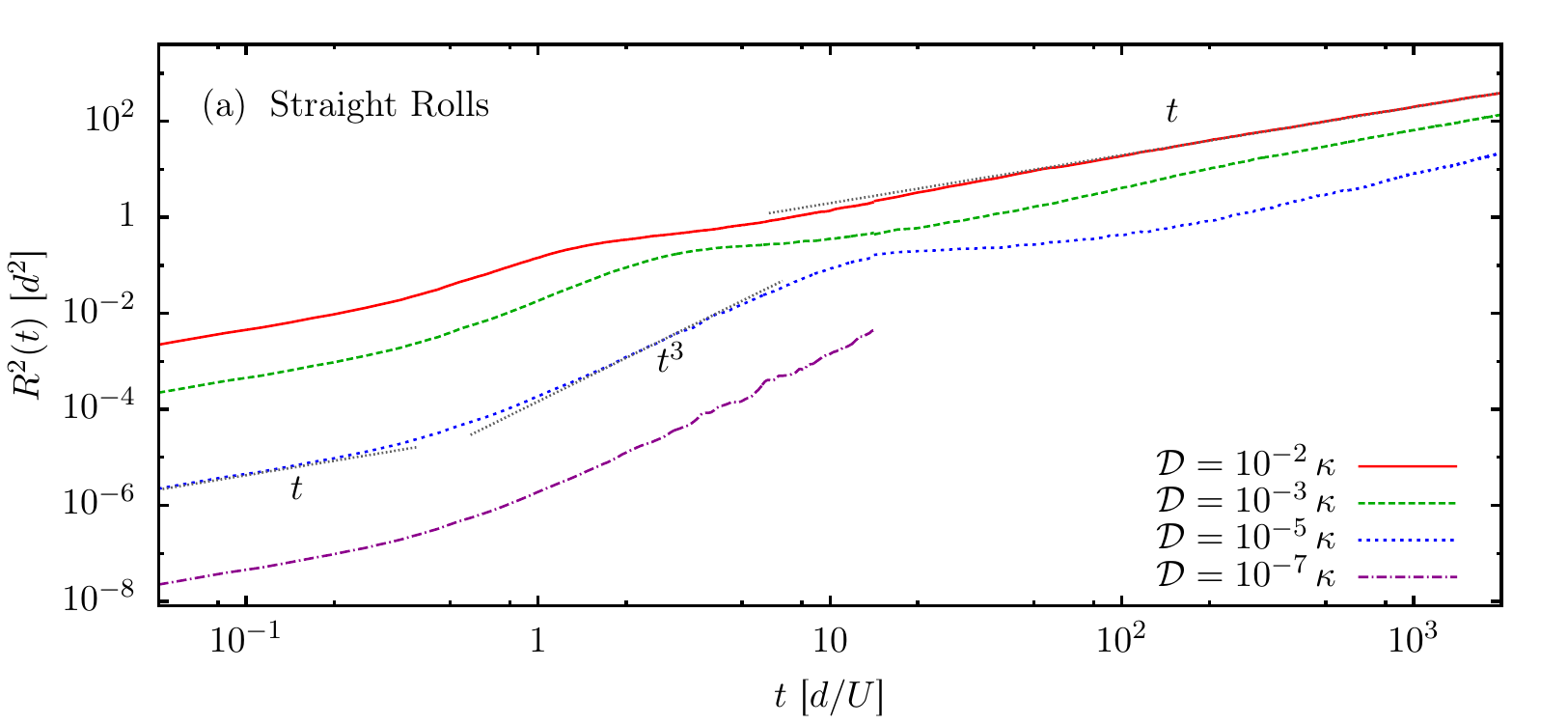}
\includegraphics[scale=0.8]{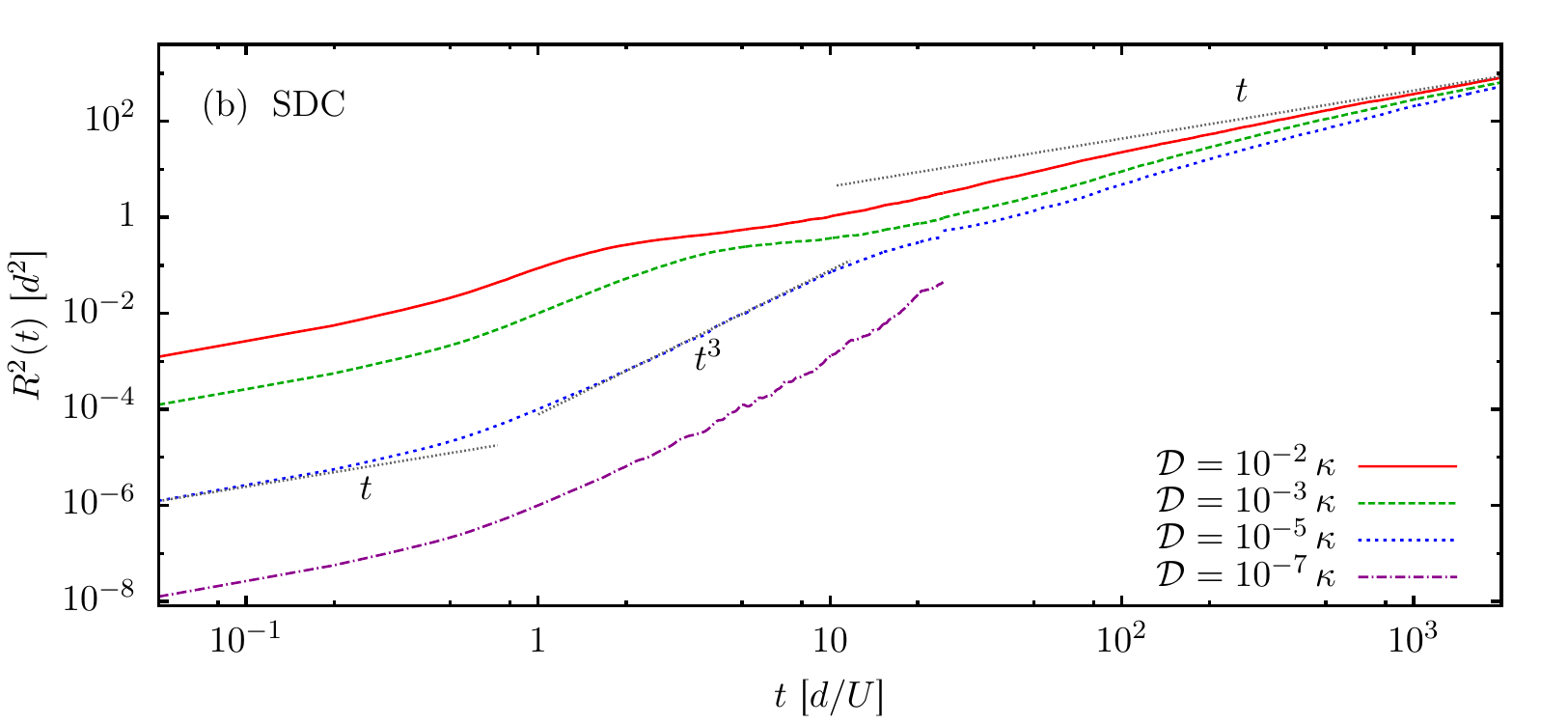}
\caption{Pair dispersion $R^2(t)$ of particles in a periodic system of size $L=200\,d$, for different particle diffusivities $\mathcal{D}$. (a) Straight rolls at $Ra=2050$ and with a characteristic flow velocity of $U=2.8d/t_v$. (b) Spiral Defect Chaos, at $Ra=2732$ with $U=4.9d/t_v$. Data averaged from 10 independent realizations of the convective system, with $10^3$ particles in each. Different data sets at a higher temporal resolution were used for $t<10t_v$ than for $t>10t_v$, hence the discontinuity in the graphs. Gray dotted lines show appropriate power laws.}
\label{fig:R2vst}
\end{figure}
We find that pair dispersion in large aspect ratio Rayleigh-\Benard convection with straight rolls and spiral defects follows three different regimes of power law behavior (Figure \ref{fig:R2vst}).
For very short times, we observe normal diffusion ($R^2\sim t$), followed by a steep Richardson-like $t^3$-dispersion up to a plateau at $R^2\approx(d/2)^2$ for intermediate times.
For very long times, we find asymptotic behavior towards a normal diffusion, but with a much higher effective diffusion coefficient.
Between the dispersion in straight rolls and Spiral Defect Chaos, we find the remarkable difference that for large times, dispersion in the chaotic state appears to have only a weak dependence on the individual particles' diffusivities.

We can consider separately the dispersion behavior for short, intermediate and long time scales.
For short times $t\ll d/U$ (Section \ref{sec:short_times}) and intermediate times $t\approx d/U$ to $R^2\approx(d/2)^2$ (Section \ref{sec:intermediate_times}), we propose an analytical model that is in good quantitative agreement with the observations (Section \ref{sec:model}).
After that, we discuss the long-term dispersion for $\sqrt{R^2}\gg d$ (Section \ref{sec:long_times}).

\subsection{Diffusive Dispersion on a Short Time Scale}\label{sec:short_times}
For short time scales much smaller than the vertical advection time ($t\ll d/U$), the particles disperse like
\begin{equation}
R^2(t)=12\mathcal{D}t,
\end{equation}
as would the case for the relative dispersion of two points that diffuse in $\mathbb{R}^3$ in the absence of advection.
This is unsurprising, as the flow field is nearly homogeneous on such small length scales and dispersion is a Lagrangian measure.

\subsection{Dispersion on Intermediate Time Scales}\label{sec:intermediate_times}
\begin{figure}[b]
\centering
\includegraphics[width=0.6\textwidth]{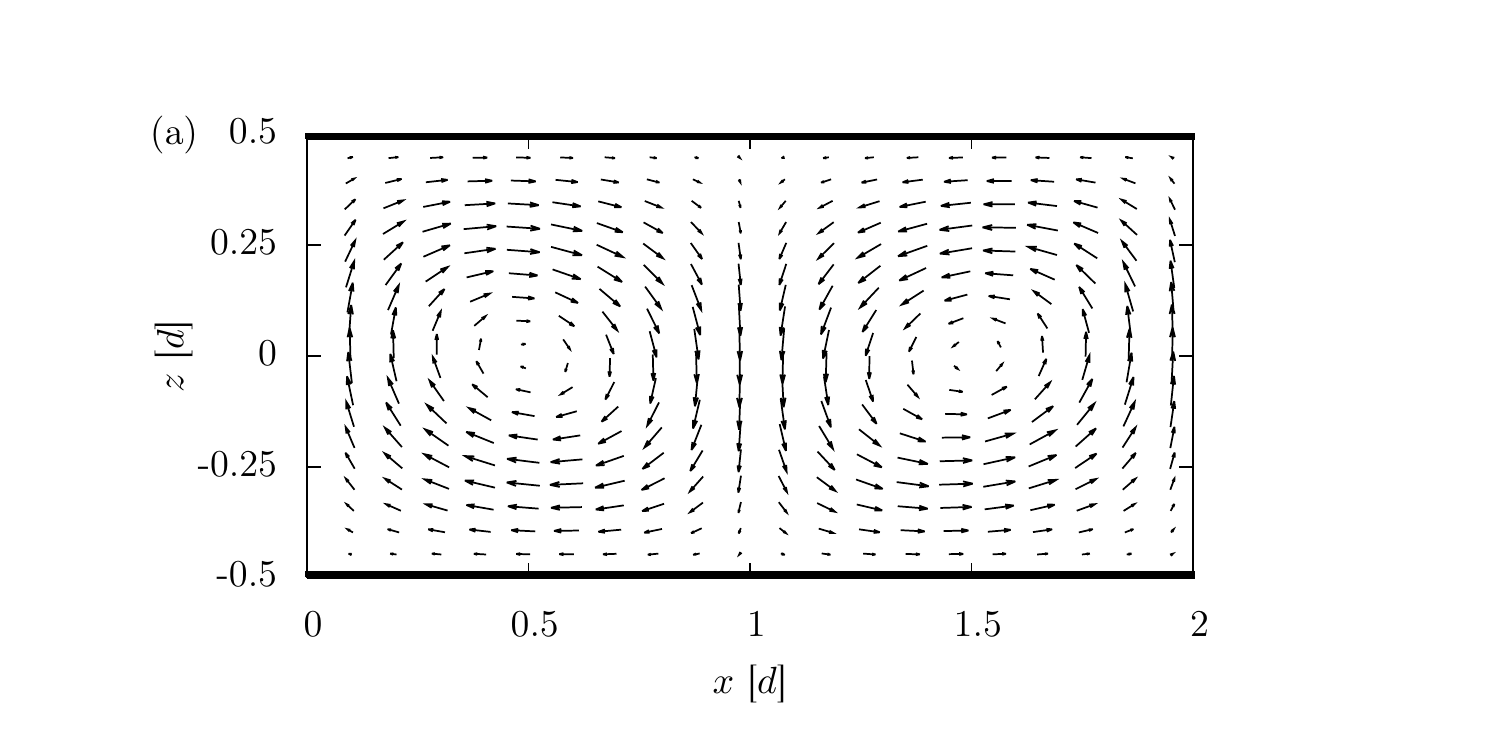}\quad
\includegraphics[width=0.3\textwidth]{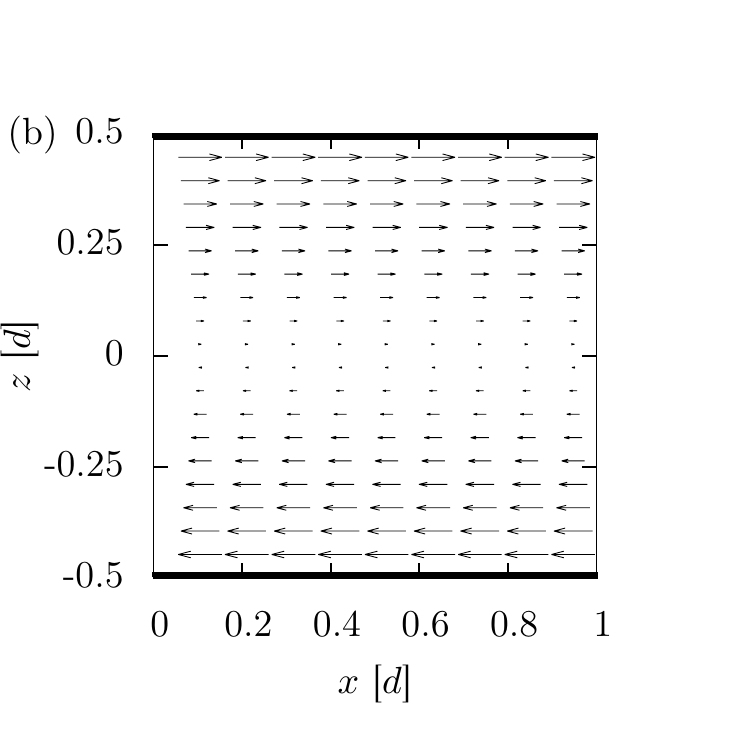}
\caption{2D cuts of velocity fields. $x$ denotes a horizontal, $z$ the vertical direction. (a) Velocity field of counter-rotating convection rolls at $Ra=2647$. (b) Velocity field in a shear flow with a constant gradient.}
\label{fig:vel}
\end{figure}
Starting from $t\approx d/U$, dispersion steepens and displays the typical Richardson-like behavior
\begin{equation}
R^2(t)\sim t^3.
\end{equation}
The $t^3$-regime starts at the same time for all particle diffusivities and in both flow configurations, and ends at dispersions around $R^2\approx(d/2)^2$.
This makes particles at very low diffusivities show the scaling over more than one order of magnitude in time.
This behavior is the same regardless of whether the flow forms straight rolls or spiral defects.
If we measure time in terms of the vertical advection time $d/U$, we observe that the dispersion data for $2050\leq Ra\leq2732$ at fixed $\mathcal{D}$ collapses in the range $0<t<2d/U$ (not shown in figures).

\subsection{Analytical Model for Dispersion on Short and Intermediate Time Scales}\label{sec:model}
The observed features can be explained by comparing the convection system to a shear flow of comparable local shear rate.
Consider a shear flow
\begin{equation}
\bi{u}=\omega z\cdot\bi{e}_x,\label{eq:cgvf}
\end{equation}
with $\omega\equiv|\partial u_x/\partial z|$ (Figure \ref{fig:vel}(b)).
In this velocity field, consider a particle initially at $\bi{x}_0=\bi{0}$ at $t_0=0$ and undergoing diffusion only in the $z$-direction.
At time $t>t_0$, its $z$-coordinate is normally distributed,
\begin{equation}
Z\sim\mathcal{N}_{0,\sigma}(z)\quad\textrm{ with }\sigma^2=2\mathcal{D}t,
\end{equation}
with $\mathcal{N}_{\mu,\sigma}$ the normal distribution of mean $\mu$ and standard deviation $\sigma$.
The $z$-position influences advection in the $x$-direction, so that the $x$-position will be distributed like
\begin{equation}
X\sim\mathcal{N}_{0,\sigma}(x)\quad\textrm{ with }\sigma^2=(2/3)\omega^2\mathcal{D}t^3
\end{equation}
This relation follows from basic properties of normal distributions, a proof is provided in \ref{sec:proof_gaussian}. For an intuitive understanding, consider that for $z=const$, $x\sim z\cdot t$. Then, for $z\sim\sqrt{t}$ due to diffusion, $x\sim t^{3/2}$ and $\sigma_x^2\sim t^3$.

Consider now a particle that diffuses in all spatial directions.
Since the velocity field is homogeneous in the $x$-direction, diffusion in that direction is independent of the $z$-position.
Thus, the position of a particle will be
\begin{eqnarray*}
X\sim\mathcal{N}_{0,\sigma_x}(x)&\quad\textrm{ with }\sigma_x^2=(2/3)\omega^2\mathcal{D}t^3+2\mathcal{D}t,\\
Y\sim\mathcal{N}_{0,\sigma_y}(y)&\quad\textrm{ with }\sigma_y^2=2\mathcal{D}t,\\
Z\sim\mathcal{N}_{0,\sigma_z}(z)&\quad\textrm{ with }\sigma_z^2=2\mathcal{D}t.
\end{eqnarray*}
The distance between two such particles with identical initial positions that diffuse independently from each other is then
\begin{eqnarray*}
\Delta X\sim\mathcal{N}_{0,\sigma_x}(\Delta x)&\quad\textrm{ with } \sigma_x^2=(4/3)\omega^2\mathcal{D}t^3+4\mathcal{D}t,\\
\Delta Y\sim\mathcal{N}_{0,\sigma_y}(\Delta y)&\quad\textrm{ with }\sigma_y^2=4\mathcal{D}t,\\
\Delta Z\sim\mathcal{N}_{0,\sigma_z}(\Delta z)&\quad\textrm{ with }\sigma_z^2=4\mathcal{D}t.
\end{eqnarray*}
The pair dispersion in a velocity field with a constant gradient is therefore
\begin{equation}
R^2(t)\equiv\langle r^2(t)\rangle=\sigma_x^2+\sigma_y^2+\sigma_z^2 =(4/3)\omega^2\mathcal{D}t^3+12\mathcal{D}t.
\end{equation}

\begin{figure}
\centering
\includegraphics[scale=0.8]{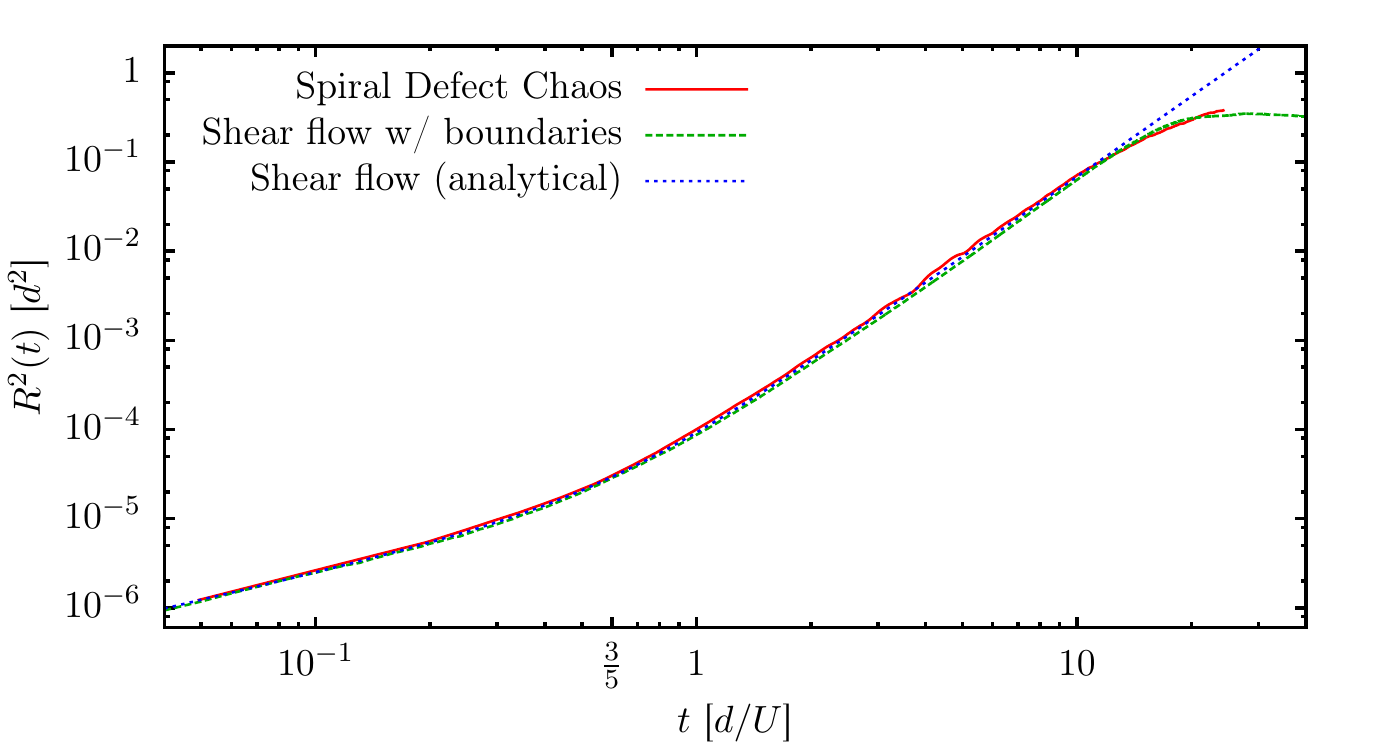}
\caption{Comparison of pair dispersion $R^2(t)$ in Rayleigh-\Benard convection with the constant gradient model. Dispersion in RBC was measured at $Ra=2732$, where $U\approx4.89\kappa/d$. Diffusivity was $\mathcal{D}=10^{-5}\kappa\approx2.04\cdot10^{-6}dU$. The velocity gradient for the shear flow model is $\omega=5U/d$. Note that the transition in scaling lies close to $t=\frac{3}{5}d/U$.}
\label{fig:R2Grad}
\end{figure}

The analytical model assumes a system that is infinitely extended, whereas particle motion in convection is confined.
To observe the effects of confinement on the model, we repeated our numerical particle tracking for Couette-type shear flow between two parallel plates, periodic along the $x$-axis with periodicity $2d$.
The resulting dispersion is shown in Figure \ref{fig:R2Grad}, where $\omega=5U/d$ was chosen as an estimate for the shear rate at the particle seed points:
Dispersion in the convection system collapses with both the numerical and analytical models for shear flow.
Due to the appropriately chosen periodicity, which prevents particles from obtaining a horizontal distance larger than $d$, the simulation of confined shear flow reproduces the plateau that we observe in the Rayleigh-\Benard convection, in addition to the scaling behavior that we predict with the analytical model.

\subsection{Effective Diffusivity on Long Time Scales}\label{sec:long_times}
\begin{figure}[!b]
\centering
\includegraphics[scale=0.8]{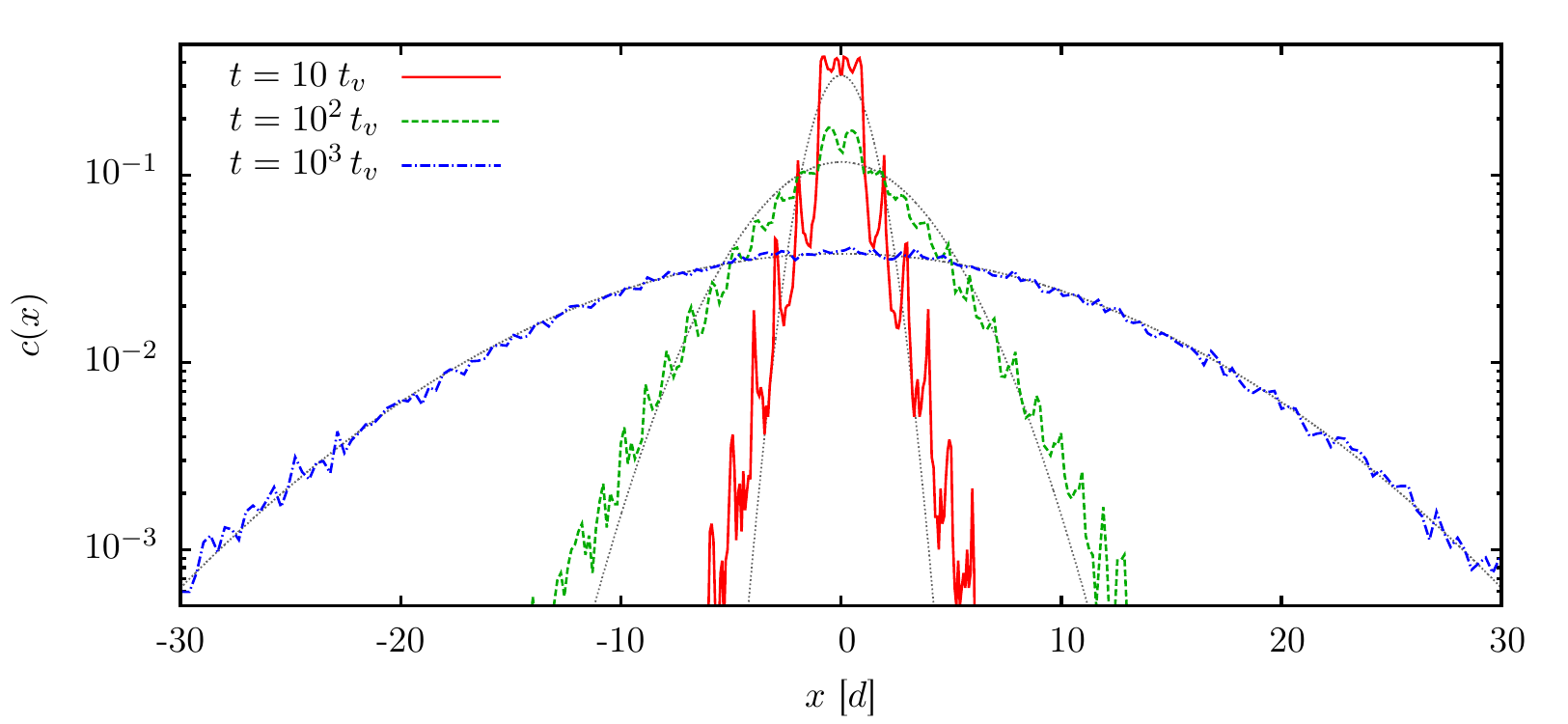}
\caption{Concentration $c$ of particles along the direction $\bi{e}_x$ perpendicular to the roll orientation, in a system of straight rolls at $Ra=2647$, for different times $t$.
Horizontal distance is given in units of the system height $d$, time in units of the vertical thermal diffusion time $t_v\equiv d^2/\kappa$.
A total number of $10^5$ particles with a diffusivity $\mathcal{D}=10^{-3}\kappa$ were seeded into the volume at a roll boundary, at midheight in the cell.
Gray dotted lines are Gaussian fits.}
\label{fig:profile}
\end{figure}

The transition to long time scales can be best visualized in a system with parallel straight rolls as shown in Figure \ref{fig:midplanetemp} (a).
Seeding a large number of diffusing particles into the system at one location allows us to study the development of their concentration with time.
We examined the spatial distribution along the direction $x$, perpendicular to the roll orientation.
The distribution assumes a Gaussian profile after a long time ($t\geq10^3\tau_v$), when the separation between the individual particles is much larger than the size of the convection rolls (Figure \ref{fig:profile}).
For shorter times, the density distribution is modulated with the spatial periodicity of the convection rolls and deviates from a Gaussian.

Since for large $t$, the pair dispersion approaches a linear increase with time ($R^2\propto t$), we define the long-term effective diffusivity in the system as
\begin{equation}
\mathcal{D}^* \equiv \lim_{t\rightarrow\infty} R^2(t)/(4t).
\end{equation}
Setting the denominator to $4t$ (instead of $6t$) takes into account that, for very long times, dispersion is dominated by the two horizontal components of the particle separations.

\begin{figure}[b]
\centering
\includegraphics[width=0.48\textwidth]{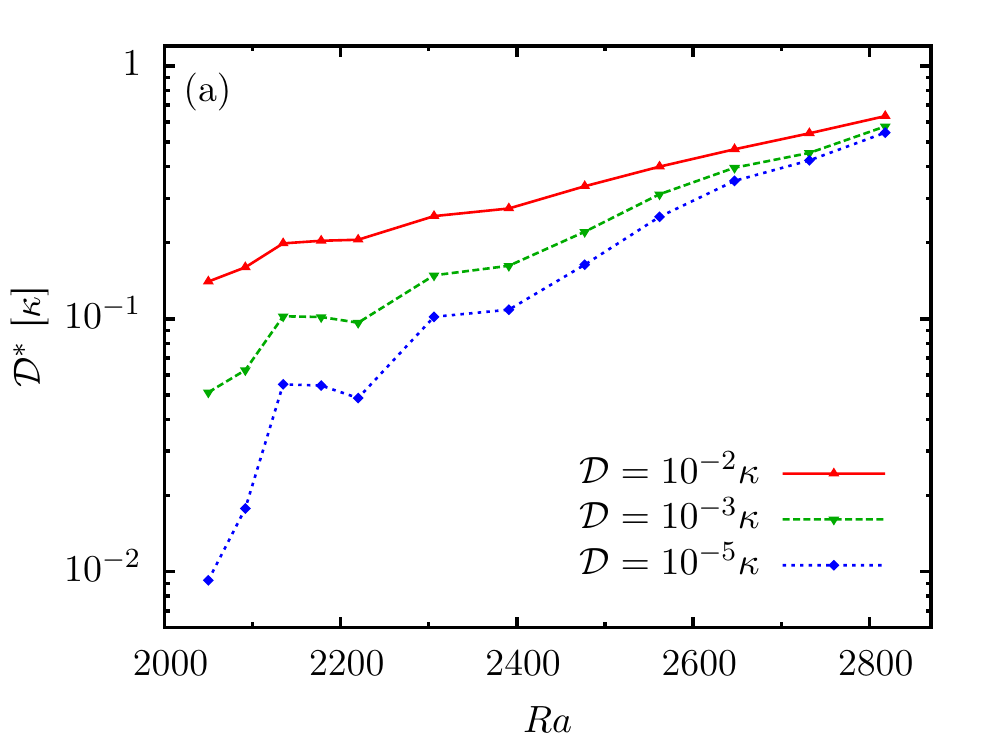}\quad
\includegraphics[width=0.48\textwidth]{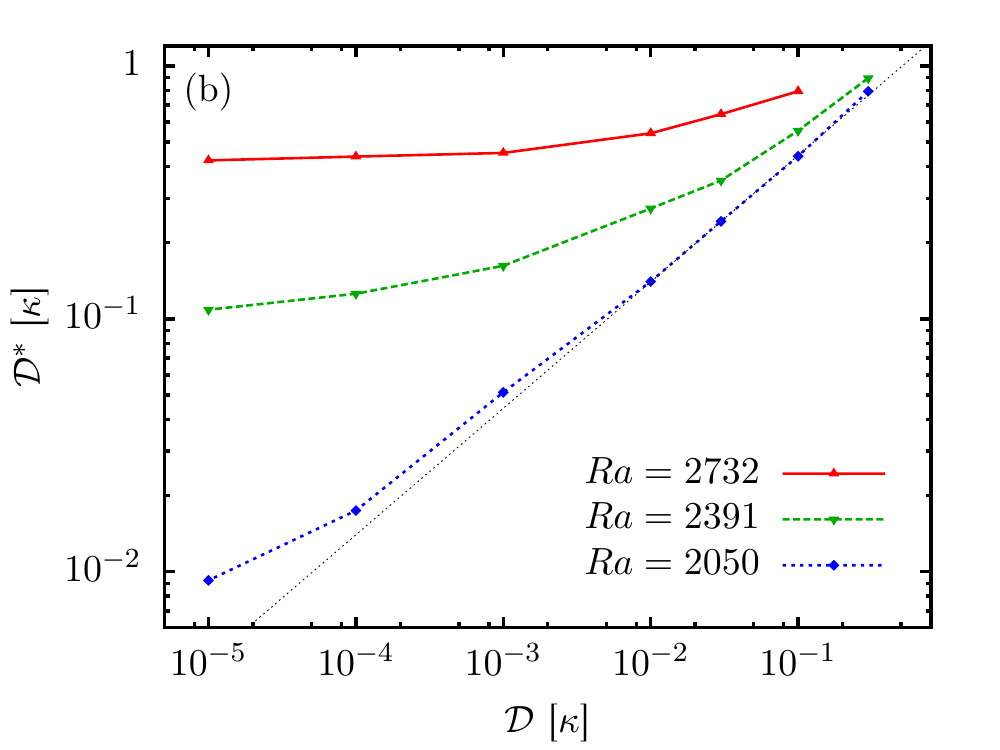}
\caption{Long-term measurements for particle dispersion.
(a) Long-term effective diffusivity $\mathcal{D}^*$, measured at $t=10^3t_v$, for systems at different Rayleigh numbers.
(b) Change of the long-term effective diffusivity $\mathcal{D}^*$ with the diffusivity $\mathcal{D}$, both normalized by the advection scale $dU$.
The gray dotted line is $\mathcal{D}^*\propto\mathcal{D}^{1/2}$.
Data averaged from 10 independent realizations of the convective system, with $10^3$ particles in each.}
\label{fig:Dstar}
\end{figure}

The so-defined effective diffusivity $\mathcal{D}^*$ exists also in systems of Spiral Defect Chaos.
It is a function of both the Rayleigh number $Ra$ of the flow, and the particle diffusivity $\mathcal{D}$ of the tracers.
The dependence on the particle diffusivity is more clear at small Rayleigh numbers, whereas higher Rayleigh numbers lead to an increase of $\mathcal{D}^*$ for all $\mathcal{D}$ (Figure \ref{fig:Dstar} (a)).
In the dispersion enhancement with $Ra$, two factors play a role:
Locally, convection is faster at a higher Rayleigh number, the dependence is roughly $U\propto\sqrt{Ra-Ra_c}$.
Additionally, the pattern of the rolls changes due to the introduction of spiral defects between $Ra\approx2000$ and $Ra\approx2300$.
A further increase in the Rayleigh number speeds up the temporal dynamics, creates more spiral cores and stronger mean currents through the system \cite{Bodenschatz.2000.AnnRevFM,Morris.1993.PRL}.

In its dependence on $\mathcal{D}$, the effective diffusivity approaches a constant $\mathcal{D}^*=const$ for small $\mathcal{D}$, and follows a power law $\mathcal{D}^*\propto\mathcal{D}^{1/2}$ for large $\mathcal{D}$ (Figure \ref{fig:Dstar} (b)).
At low particle diffusivities, the effective particle transport is dominated by the pattern dynamics and much faster for very active Spiral Defect Chaos, whereas at a high particle diffusivities, we find $\mathcal{D}^*\sim\mathcal{D}^{1/2}$ regardless of the underlying flow pattern.
The transition point depends strongly on the Rayleigh number.

We note that this effect cannot be attributed to the effects of a generally faster advection speed $U$ at higher $Ra$, but rather the additional effect of the changes in the convection pattern:
The general behavior of $\mathcal{D}^*$ remains unchanged when we normalize diffusivities by the advection velocities (i.e. express $\mathcal{D}$ via the P\'eclet number $\mathcal{P}\equiv dU/\mathcal{D}$).

\section{Discussion}

\subsection{Dispersion on Short and Intermediate Time Scales}
We have resolved particle dispersion in Spiral Defect Chaos for very short times.
On these time scales, we find that dispersion is accelerated by the shear in the flow field.
Particle dispersion scales with $t^3$, which leads to a quick mixing of the diffusing particles throughout a convection roll.

Diffusion in a shear flow reproduces all characteristics of the short-term behavior of dispersion in the convective system:
For very small times, we find 3D normal diffusion with
\begin{equation*}
R^2(t)=12\mathcal{D}t.
\end{equation*}
The transition to the Richardson-like $t^3$-scaling occurs when the two terms $12\mathcal{D}t$ and $(4/3)\omega^2\mathcal{D}t^3$ have the same order of magnitude, i.e. around $t = 3/\omega$.
This transition point only depends on the vorticity $\omega$ and is independent of the particle diffusivity.
Afterwards, the $t^3$-term dominates, and we find
\begin{equation*}
R^2(t)=(4/3)\omega^2\mathcal{D}t^3.
\end{equation*}
Due to the presence of top and bottom boundaries, the steep increase in dispersion is limited.
When the separation distance of two particles is on the order of $d$, the velocity field cannot be modeled by a shear flow anymore, and the fast dispersion mechanism breaks down.

Recently, Benveniste and Drivas\cite{Benveniste.2014.PhysRevE} have investigated backwards dispersion of diffusing particles in turbulent flow.
Applying their calculation to the shear flow $\bi{u}=\omega z\cdot\bi{e}_x$ (Eq. \ref{eq:cgvf}) at viscosity $\nu$ predicts a backwards dispersion of
\begin{equation}
R_b^2(t)~=~(4/3)\langle\epsilon\rangle t^3
\end{equation}
for particles of diffusivity $\mathcal{D}=\nu$, with a mean dissipation $\langle\epsilon\rangle=\nu\omega^2$.
Since the simple shear flow is independent of viscosity, this result for backwards dispersion can be applied to any diffusivity and matches our description for the forward dispersion at intermediate times.

\subsection{Dispersion on Long Time Scales}
When pair separation reaches the length scale of the system height, the transition between the rolls becomes relevant for transport, a model for which has been suggested by Shraiman \cite{Shraiman.1987.PhysRevA}.
In accordance to that model, impurities in a Rayleigh-\Benard pattern of parallel, straight convection rolls have been found to spread throughout the rolls like in a normal diffusive process, albeit with an increased effective diffusion \cite{Solomon.1988.PhysFluids,Gollub.1989.PhysScripta}.
We find that this is also true for particles in Spiral Defect Chaos, where rolls are not stationary and the flow field is truly three-dimensional.

For small particle diffusivities, the effective diffusivity goes like $\mathcal{D}^*\sim\mathcal{D}^{1/2}$, as has been observed numerically and experimentally by Gollub and Solomon\cite{Gollub.1989.PhysScripta} in a system of stationary convection rolls.
When increasing the particle diffusivity, the dependence of the effective diffusivity on the particles' diffusivities vanishes, and the dispersion appears characteristic to the flow pattern itself.
Our results complement the findings of Chiam \etal\cite{Chiam.2005.PRE}, who studied the enhancement of scalar diffusion in a similar system of Spiral Defect Chaos in terms of the P\'eclet number $\mathcal{P}$, and found a transition in scaling behavior at $\mathcal{P}\approx10^2$ for $Ra=3074$ and higher.
This transition hints towards a two-fold effect of the stochastic fluid motion on particle dispersion:
One part that dominates when particle diffusivity is small, and disperses particles purely by advecting them; and a second part that results from the interplay of advection and particle diffusion, which dominates at higher particle diffusivities (corresponding to lower P\'eclet numbers).

\section{Conclusion}
We found that the Richardson-like $t^3$ scaling of pair dispersion known from fully developed turbulence also occurs in laminar thermal convection flow, if the effects of the stochastic turbulent motion are replaced by a particle diffusivity $\mathcal{D}$.
Unlike in homogeneous and isotropic turbulence, the size of the vortex structures in Rayleigh-\Benard convection is fixed, creating an upper bound to the quick vertical dispersion.

Due to the strong difference in the transport mechanisms before and after the transition from 3D motion to effectively 2D motion, this transition can be clearly observed in the dispersion as a plateau.
If the large-scale flow pattern shows strong enough variations in time, we find that dispersion loses its dependence on the local details of stochastic motion, and depends only on the orientation pattern of the convection rolls, the correlation distance of which is small in Spiral Defect Chaos (i.e. on the order of a few roll diameters).
This strong cutoff in correlation length is specific to the vertically restricted systems and does not occur in fully developed 3D turbulence.

It remains an open question how exactly the long-term dispersion can be described in terms of the pattern dynamics and correlation properties of the flow pattern, and whether a $\mathcal{D}$-independent effective diffusivity $\mathcal{D}^*$ can also be found at lower Rayleigh numbers, when only few dislocations are present.
Since larger correlation lengths in the pattern require longer observations and larger system sizes, the simulation efforts are very high for these systems.

\section*{Acknowledgments}
We thank Haitao Xu for helpful comments. We are in debt to Jens Zudrop and Werner Pesch for their very significant help with the pseudo-spectral code.

\clearpage
\appendix\addtocontents{toc}{\protect\setcounter{tocdepth}{0}}

% Supplementary material
\section{Details of the Numerical Solver}\label{sec:solver_details}
The numerical solver to calculate the flow field $\bi{u}(\bi{x},t)$ solves the Boussinesq equations for flow velocity and temperature, which are
\begin{eqnarray}
Pr^{-1}\left(\frac{\partial\bi{u}}{\partial t}+\bi{u}\cdot\nabla\bi{u}\right)-\nabla^2\bi{u}+\nabla p = Ra\,T\bi{e}_z,\\
\nabla\cdot\bi{u} = 0,\\
\frac{\partial T}{\partial t} - \bi{u}\cdot\bi{e}_z + \bi{u}\cdot\nabla T = \nabla^2 T,
\end{eqnarray}
where $\bi{e}_z$ is a unit vector in the (upward) vertical direction and $T$ the deviation of the local temperature from a linear temperature profile between the top and bottom plate.
The solenoidal velocity field is decomposed into its toroidal part $f$ and poloidal part $g$, and mean flow components $F(z,t)$, $G(z,t)$ in the two orthogonal horizontal directions, such that
\begin{equation}
\bi{u} = \left(\begin{array}{c}\partial_x\partial_z\\\partial_y\partial_z\\-\partial_x^2-\partial_y^2  \end{array}\right) f + \left(\begin{array}{c}\partial_y\\-\partial_x\\0 \end{array}\right) g + \left(\begin{array}{c}F\\G\\0 \end{array}\right).
\end{equation}

We assume a periodic domain in the horizontal directions and no-slip boundary conditions at the top and bottom plates. In the horizontal directions, the toroidal and poloidal velocity potentials ($f$ and $g$) and the temperature $T$ can then be written as Fourier series. To account for the different boundary conditions in the vertical $z$-direction, the $z$-dependence of $f$, $T$, $F$ and $G$ is expressed as a series of sine functions. The $z$-dependence of the poloidal velocity potential $g$ is expressed as a series of Chandrasekhar functions. This spectral representation is used to construct the weak-form Galerkin solution of the equation system.

To solve for the time development of the system, we use an implicit-explicit pseudospectral scheme. We decompose the equations into their linear and nonlinear parts. The nonlinearity is evaluated explicitly and in real space. The linear part of the equation system, due to the orthogonality of the horizontal Fourier basis, can be expressed in terms of a block diagonal matrix. The vertical basis is not orthogonal, but the low Rayleigh number allows us to keep the number $N_z$ of vertical basis functions low (values between 2 and 5 were used). Both the block size of the linear matrix, and the number of Fourier transforms for the nonlinear part scale with $N_z$. We add the nonlinearity in an explicit two-step Adams-Bashforth method and then use the linear matrix for an implicit Euler-step.

The simulation code is parallelized to run on a GPU. Unless stated otherwise, we use a resolution of $1024\times1024$ modes in the horizontal and 2 modes in the vertical direction to simulate a system of aspect ratio $\Gamma=200$ with a time step of $\rmd t=10^{-2}$.

Advection of the tracer particles is performed by sampling the velocity field in a regular grid and using trilinear interpolation to get the fluid velocity at a particle's position. In the horizontal directions, the number of sampling points is chosen equal to the number of modes, while vertically, 128 sampling points are chosen. Particles are advected in a 4th-order Runge-Kutta scheme. The final position after each time step is modified by adding a Gaussian-distributed random variable with a standard deviation of $\sigma=\sqrt{2\mathcal{D}\,\rmd t}$ to each position component. Random variables are drawn from independent random number generators for the spatial directions and the different particles. Pseudorandom number generation uses the XORWOW algorithm.

\section{Quantitative Description of Particle Diffusion in a Shear Flow}\label{sec:proof_gaussian}
We will show that in a shear flow $\bi{u}=\omega z\cdot\bi{e}_x$, diffusion of a particle in the direction $\bi{e}_z$ will at time $t$ lead to a probability distribution along the $x$-axis of
\begin{equation}
X \sim \mathcal{N}_{0,\sigma}(x)\quad\textrm{ with }\sigma^2=(2/3)\omega^2\mathcal{D}t^3,
\end{equation}
with $\mathcal{N}_{\mu,\sigma}$ the normal distribution of mean $\mu$ and standard deviation $\sigma$.

For simplicity, let motion be restricted to the $x$-$z$-plane. We make use of the fact that the individual steps of the diffusive motion can be considered discrete (but the limit $\textrm{d}t\rightarrow0$ exists). Consider a single particle that starts at position $\bi{x}_0=\bi{0}$ and moves $N$ steps in time $t$, so that each time step is $\textrm{d}t\equiv t/N$. Let $\textrm{d}z_n$ be the $n$'th position increment of its trajectory, such that its $z$-position at time step $m$ is
\begin{equation}
 z_m~=~\sum_{n=1}^m\textrm{d}z_n.\label{eq:znz}
\end{equation}
Let $\textrm{d}z_n\sim\mathcal{N}_{0,\sigma_z}(x)$ with $\sigma_z^2=2\mathcal{D}t/N$. The particle's velocity depends only on its $z$-position, so that its velocity in the $x$-direction at time step $m$ is
\begin{equation}
 v_m = \omega z_m = \omega\sum_{n=1}^m\textrm{d}z_n
\end{equation}
and the particle's final $x$-position can be written as a discrete integral over those velocities. At time step $N$, the $x$-position is
\begin{eqnarray}
 x_N &= \sum_{m=1}^Nv_m\cdot\textrm{d}t \nonumber\\
 &= \sum_{n=1}^N (\omega t n/N)\cdot\textrm{d}z_{(N+1-n)}.
\end{eqnarray}
Since all random position increments $\textrm{d}z_n$ are from the same distribution, we write
\begin{equation}
 x_N = \sum_{n=1}^N \left(\omega tn/N\right)\textrm{d}z_n.
\end{equation}

For independent variables $X_n$ that have normal distributions of mean $\mu_n$ and variance $\sigma_n^2$, we know that $Y\equiv \sum_n c_nX_n$ is normal distributed with mean $\mu=\sum_nc_n\mu_n$ and variance $\sigma^2=\sum_nc_n^2\sigma_n^2$. Here, we have $c_n=\omega tn/N$, $\mu_n=0$ and $\sigma_n^2=2\mathcal{D}t/N$. Hence, the position $x_N$ has the distribution
\begin{equation}
X_N \sim \mathcal{N}_{0,\sigma}(x)\quad\textrm{ with }\sigma^2=\sum_{n=1}^N\frac{2\omega^2\mathcal{D}t^3n^2}{N^3}.
\end{equation}
The limit $N\rightarrow\infty$ exists, and yields the result
\begin{equation}
 \lim_{N\rightarrow\infty}\sigma^2 = \lim_{N\rightarrow\infty}\sum_{n=1}^N\frac{2\omega^2\mathcal{D}t^3n^2}{N^3} 
 = (2/3)\omega^2\mathcal{D}t^3.
\end{equation}
In the limit of small steps, we therefore arrive at the final result
\begin{equation}
X \sim \mathcal{N}_{0,\sigma}(x)\quad\textrm{ with }\sigma^2=(2/3)\omega^2\mathcal{D}t^3.
\end{equation}

Since the velocity field is homogeneous, shifting the initial position $\bi{x}_0$ away from the origin does not change the variance of the resulting Gaussian distribution, but moves the mean only. When we investigate the Lagrangian pair dispersion, we consider the relative position between two particles. As they are starting from the same initial position, the mean value of their relative separation is zero.

\end{document}